\documentclass[%
 reprint,
 amsmath,amssymb,
 aps,
]{revtex4-2}

\usepackage{graphicx}
\usepackage{dcolumn}
\usepackage{bm}
\usepackage{lineno}
\usepackage{adjustbox}
\usepackage{epsfig}
\usepackage{epstopdf}
\usepackage{float}
\usepackage{amssymb}
\usepackage{color}
\usepackage{bm}
\usepackage{amsfonts}
\usepackage{lineno,hyperref}
\usepackage{array}
\usepackage{microtype}
\usepackage{xcolor}
\usepackage{caption}
\usepackage{subcaption}

\begin{document}

\preprint{APS/123-QED}

\title{Observational constraints on the Emergent Universe with interacting non-linear fluids and its stability analysis}

\author{Anirban Chanda $^1$}
\email{aniphys93@nbu.ac.in}
\author{Bikash Chandra Roy $^1$}%
 \email{bcroy.bcr25@gmail.com}
\author{Kazuharu Bamba $^2$}
\email{bamba@sss.fukushima-u.ac.jp}
\author{Bikash Chandra Paul $^1$}
\email{bcpaul@nbu.ac.in}
\affiliation{$^1$Department of Physics, University of North Bengal, Raja Rammohunpur, 734013, India\\
$^2$Faculty of Symbiotic Systems Science, Fukushima University, Fukushima 960-1296, Japan}

%

\date{\today}

\begin{abstract}
We investigate a flat Emergent Universe (EU) with a nonlinear equation of state which is equivalent to three different compositions of fluids. In the EU, initially, the evolution of the universe began with no interaction, but as time evolves, an interaction sets in among the three fluids leading to the observed universe. The characteristic of an EU is that it is a singularity-free universe that evolves with all the basic features of the early evolution. A given nonlinear equation of state parameter permits a universe with three different fluids. We get a universe with dark energy, cosmic string, and radiation domination to begin with, which at a later epoch transits into a universe with three different fluids with matter domination, dark matter, and dark energy for a given interaction strength among the cosmic fluids. Later the model parameters are constrained using the observed Hubble data and Type Ia Supernova (SnIa) data from the Pantheon data set. The classical stability analysis of the model is performed using the square speed of sound. It is found that a theoretically stable cosmological model can be obtained in this case, however, the model becomes classically unstable at the present epoch when the observational bounds on the model parameters are taken into account.\\
\\
{\it {\bf Key Words :} Emergent Universe, Observational Constraints, Cosmological Parameters, Classical Stability}
\end{abstract}

\maketitle

\section{Introduction}
\label{sec:intro}

Modern cosmology in the present decade is based on astronomical observations. We witnessed a transition from speculative science to experimental science because of precision measurements from different cosmological missions. The observations predict that the universe is not only expanding but is accelerating \cite{riess_observational_1998}, \cite{1999ApJ...517..565P}, \cite{Bennett_2003}, \cite{riess2004type}, \cite{eisenstein_detection_2005}, \cite{bennett_nine-year_2013}. After the discovery of cosmic microwave background radiation (CMBR), big-bang cosmology became the standard model to study the evolution of the universe having a beginning at some finite past. However, the standard model of cosmology is plagued with a number of problems, namely, the horizon problem, flatness problem, singularity problem etc. \cite{brandenberger_inflationary_1999}, \cite{kolb_early_2019}. To resolve these problems, it has been proposed that in the early stage of evolution of the universe, a rapid expansion of space took place which is known as cosmic inflation. A homogeneous scalar field in the framework of standard cosmology permits such an inflation \cite{Guth1981}, \cite{sato_first-order_1981}, \cite{linde2002inflation}. Furthermore, inflation can address the large-scale structure formation of the universe. The present observational data predict that the present universe is passing through a phase of cosmic acceleration. This late time accelerating phase of the universe may be explained in the standard model by adding a positive cosmological constant $(\Lambda)$ in Einstein’s field equations (EFE). The $\Lambda$ cold dark matter ($\Lambda CDM$) model is currently the most favored model in cosmology which matches well with astronomical observations. The $\Lambda CDM$ model is found to have some conceptual issues, namely, the exact nature of its main constituents is not yet known. There are issues like fine-tuning and cosmic coincidence which are to be resolved \cite{sahni_case_2000}, \cite{carroll_cosmological_2001}, \cite{padmanabhan_cosmological_2003}, \cite{peebles_cosmological_2003}. Recently it is observed from the CMB measurements that the expansion rate of the universe based on local data is different in comparison to the expansion rate that the universe had in the past \cite{freedman_carnegie-chicago_2019}, \cite{yuan_consistent_2019}, \cite{freedman_measurements_2021}, \cite{riess_cosmic_2021}, \cite{soltis_parallax_2021}. This issue is known as the Hubble tension.\\
Since General relativity (GR) and normal matter cannot support the present acceleration of the universe, an alternative is to modify the gravitational or matter sector of the EFE. Modifications in the matter sector lead to different dynamical DE models, namely, Chaplygin gas \cite{Kamenshchik_2001} and its variations \cite{Bento_2002}, \cite{Benaoum_2002}, models consisting of one or more scalar fields namely, quintessence \cite{chiba_quintessence_1999}, \cite{amendola_coupled_2000}, \cite{martin_quintessence_2008}, etc. A detailed review on different DE models including quintessence, K-essence, Tachyon, Pantom etc. can be found in the Refs. \cite{COPELAND_2006,Padmanabhan_2007,Durrer_2007,bamba2012dark}. On the other hand modifications in the gravitational sector led to the proposal of different modified theories, namely, $f(R)$ theories of gravity \cite{sotiriou_fr_2010}, \cite{De_Felice_2010}, $f(R,T)$ gravity \cite{PhysRevD.84.024020} with $T$ being the trace of the energy-momentum tensor, modified Gauss-Bonnet gravity \cite{NOJIRI20051}, $f(\mathcal{T})$ gravity \cite{ferraro_modified_2007}, \cite{ferraro_born-infeld_2008}, \cite{bengochea_dark_2009} where $\mathcal{T}$ is the torsion scalar, $f(Q)$ gravity \cite{Jim_nez_2018} where $Q$ is the non-metricity scalar, Brane world gravity \cite{Maartens_2010}, Horava-Lifshitz theory of gravity (\cite{wang_horava_2017}), etc. The modified theories of gravity are tested for the unification of the early inflationary phase with the late time acceleration phase \cite{nojiri_modified_2006}. In the literature different modified gravitational theories are considered to explain several astrophysical and cosmological phenomena and the viability of these models is also tested using astronomical observations \cite{starobinsky198030}, \cite{RUDRA2021115428}, \cite{PhysRevD.76.044027}, \cite{mandal2020cosmography}.\\
Cosmological models which are free from the initial singularity, have no horizon problem, and no quantum gravity (QG) regime are promising in this context. The "Emergent Universe" (EU) scenario proposed by Ellis and Maartens is one such model which can avoid the singularity problem of Big Bang cosmology \cite{Ellis2003}. In the EU scenario, the universe emerges as an Einstein static universe in the infinite past ($t \rightarrow - \infty$) and avoids the initial singularity by staying large at all times. The universe gradually expands slowly to attain a Big Bang phase of expansion. In the EU model, an inflationary universe emerges from a static phase and eventually leads to a macroscopic universe that occupies the present observed universe in its entirety. Once inflation starts it remains in that phase which can provide an explanation for the present acceleration. Ellis et al. \cite{Ellis_2003} obtained an EU scenario for a closed ($k=1$) universe considering a minimally coupled scalar field ($\phi)$ with a special choice of potential where the universe exits from its inflationary phase followed by reheating when the scalar field starts oscillating around the minimum of the potential. Later it was shown that such a potential occurs naturally by the conformal transformation of the Einstein-Hilbert action with $\alpha R^2$ term, where $\alpha$ is a coupling constant. Present observations predict that the universe is flat having almost zero spatial curvature. EU scenario in a flat universe can be obtained in a semi-classical theory of gravity. It is also shown that in Starobinsky model, EU model can be obtained considering a flat Robertson–Walker (RW) spacetime geometry with all its features \cite{mukherjee2005emergent}. Another interesting class of EU model in the standard GR framework was proposed by Mukherjee et al. \cite{Mukherjee_2006} considering a non-linear equation of state (nEoS) in a flat universe. In this framework, the cosmic fluid is equivalent to a mixture of normal and two different fluids, one of them is an exotic kind  described by a nEoS which is: 
\begin{equation}\label{eueos}
p = A\rho - B\sqrt{\rho},    
\end{equation}
where $A$ and $B$ are constant parameters. The composition of cosmic fluid is determined for a given value of the parameter $A$. The EU models are explored in different theories of gravity namely, Brans–Dicke theory \cite{del2007emergent}, brane world cosmology \cite{Banerjee2008}, Gauss–Bonnet modified gravity \cite{paul2010emergent}, Loop quantum cosmology \cite{mulryne2005emergent}, Energy-momentum squared gravity \cite{khodadi2022emergent}, $f(R,T)$ gravity \cite{debnath_observational_2020}, etc. Beesham ${\it et. al.}$\cite{beesham_emergent_2009-1} studied the EU model using a non-linear sigma model. An EU model with particle creation and irreversible matter creation is studied by Ghosh and Gangopadhyay using a thermodynamical approach \cite{Ghosh_2017}. The validity of EU models is studied using recent cosmological observations with the estimation of the observational constraints on the model parameters \cite{paul_constraints_2010-1}, \cite{paul_emergent_2011}, \cite{ghose_observational_2012-1}. Recently \cite{paul_observational_2019} studied the EU scenario considering cosmic fluids permitted by nEoS in addition to viscosity. The above model determines the observational bounds of the model parameters. In the present work, we investigate the effect of interaction present among the components of the cosmic fluid to estimate the bounds on the model parameters of an EU. In the original EU model \cite{Mukherjee_2006}, the composition of the cosmic fluid is fixed, to begin with, and it cannot explain satisfactorily the different phases of the evolutionary history of the universe. For an EU with radiation domination to begin with, the other two constituents namely cosmic string and DE contribute insignificantly to the total energy density in the early universe for $A=\frac{1}{3}$. As the interaction sets in, the EU transits from a radiation-dominated phase to a matter and DE-dominated phase. The observational bounds are determined for the late universe using the current Observed Hubble Dataset (OHD) \cite{Sharov_2018} as well as Pantheon supernova compilation \cite{Scolnic_2018} and it is found that the analysis differs significantly from the early studies.\\
In the literature, a class of cosmological models where the evolution of the cosmic fluids are probed with interaction and energy exchange from one sector of the fluid to the other are specially interesting. Recently, different interacting cosmological models with interaction among the dark sectors gained popularity because the present universe is not only expanding it is accelerating and we do not have a definite theory to explain the feature \cite{PhysRevD.73.103520}, \cite{PhysRevD.81.043525}, \cite{PhysRevD.81.023007}, \cite{PhysRevD.83.023528}, \cite{PhysRevD.85.107302}. Such an interacting scenario in the evolution of the universe is found in M theory \cite{banks2001mtheory} and inflationary models \cite{starobinskii_spectrum_1979}, \cite{billyard_interactions_2000}. The energy conservation equation is violated by the individual fluid components in the case of interacting cosmology, however, the total energy density remains conserved. In the present work, we consider interaction among the cosmic fluids which play a crucial role in developing a consistent cosmological model. Interacting cosmology can provide a reasonable explanation of the cosmic coincidence problem. It is well known that there is an explicit tension between the cosmological measurements made using the data from the early and late universe. Specifically the tension in $H_{0}$ and $S_{8}$ are of particular importance. The interacting cosmic fluid scenario can alleviate these tensions up to a certain degree \cite{amendola_consequences_2007}, \cite{mohseni_sadjadi_thermodynamics_2007}, \cite{quartin_dark_2008}, \cite{PhysRevD.81.043525}, \cite{10.1111/j.1365-2966.2009.16115.x}, \cite{di_valentino_can_2017-1}, \cite{kumar_echo_2017}, \cite{pan_astronomical_2018}. The motivation of our work is to explore the EU scenario which may evolve from a radiation epoch to a matter and DE dominated epoch in the presence of interaction among the fluids that sets in at a late time.\\
The paper is organized as: In sec. (\ref{fieldequ}), the basic field equations for the EU are given. In sec. (\ref{intf}), we introduce interaction among the cosmic fluids that sets in at time $t>t_{i}$ and the conservation equations for the fluid components are rewritten. The effective EoS parameters in the presence of interaction determined by the strength of the interaction are obtained. In sec. (\ref{obssec}) we use the observational data sets, namely, the OHD and the recent Pantheon compilation of 1048 Type Ia Supernovae (SnIa), to constrain the model parameters. The statistical inferences for the EU model are studied by the determination of the Akaike Information Criterion (AIC) and Bayesian Information Criterion (BIC), which are shown in sec (\ref{aic}). Using the Markov Chain Monte Carlo (MCMC) methods, cosmological parameters and the classical stability of the model are explored in sec. (\ref{cosmo}). Finally, the results obtained in the analysis are summarized in sec. (\ref{res}) followed by a brief discussion.
\section{Field Equations}
\label{fieldequ}

We consider a spatially flat, homogeneous, and isotropic spacetime described by the Robertson-Walker (RW) metric, which is given by,
\begin{equation}\label{eq1}
    ds^{2} = -dt^{2}+a^{2}(t)\Big[dr^{2}+r^{2}(d\theta^{2}+\sin^{2}\theta\; d\phi^{2})\Big],
\end{equation}
where $a(t)$ is the scale factor of the universe and $r$, $\theta$, and $\phi$ are the dimensionless comoving coordinates.\\
The Einstein field equation (EFE) is given by,
\begin{equation}\label{eq2}
    R_{\mu\nu}-\frac{1}{2}g_{\mu\nu}R=8\pi G \; T_{\mu\nu},
\end{equation}
where, $R_{\mu\nu}$ is the Ricci tensor, $R$ is the Ricci scalar, $g_{\mu\nu}$ is the metric tensor and $T_{\mu\nu}$ is the energy-momentum tensor of the cosmic fluid. Using the RW metric given by eq. (\ref{eq1}), the time-time and the space-space components of the EFE become,
\begin{equation}\label{eq3}
    \rho=3\Big(\frac{\dot{a}^{2}}{a^{2}}\Big),
\end{equation}
\begin{equation}\label{p}
    p = -\left[2\frac{\ddot{a}}{a}+\left(\frac{\dot{a}}{a}\right)^{2}\right],
\end{equation}
where, $\rho$ denotes the energy density of the cosmic fluid, $p$ denotes the pressure and we have assumed natural units $i.e.$, $c=1$ and $8\pi G=1$. The conservation equation is,
\begin{equation}\label{eq4}
    \dot{\rho}+3H(\rho+p)=0,
\end{equation}
where, $H=\frac{\dot{a}}{a}$ is the Hubble parameter and $p$ is the pressure of the fluid. 

\subsection{Physical analysis of Emergent Universe}
Using eqs. \ref{eq3} and \ref{p} in eq. \ref{eueos} we arrive at a second-order differential equation for the scale factor given by,
\begin{equation}\label{scale1}
    2\frac{\ddot{a}}{a}+(3A+1)\left(\frac{\dot{a}}{a}\right)^{2}-\sqrt{3}B\frac{\dot{a}}{a}=0.
\end{equation}
Integrating the above equation twice we obtain the scale factor ($a(t)$) which is given by, 
\begin{equation}\label{a}
    a(t) = \Big[\frac{3K(1+A)}{2}\left(K_{1}+\frac{2}{\sqrt{3}B}e^{\frac{\sqrt{3}Bt}{2}}\right)\Big]^{\frac{2}{3(1+A)}},
\end{equation}
where $K$ and $K_{1}$ are two integration constants. It is evident that if $B<0$ it leads to a singular universe, and if $B>0$ and $A>-1$ one gets a non-singular solution. The latter solution is interesting and used to obtain an EU. The scale factor $a(t)$ remains finite even at infinite past ($t\rightarrow -\infty$). Thus the universe emerged from an initial Einstein static phase in this scenario.\\ 
Using eq.(\ref{eueos}) and eq.(\ref{eq4}) we obtain the energy density which is given by,
\begin{equation}\label{eq5}
\rho = \frac{B^{2}}{(1+A)^{2}} + \frac{2BK}{(1+A)^{2}}\frac{1}{a^{\frac{3}{2}(1+A)}} + \frac{K^{2}}{(1+A)^{2}}\frac{1}{a^{3(1+A)}}.
\end{equation}
It is evident that there are three terms in the energy density $(\rho_{1}, \rho_{2}, \rho_{3})$. Now we obtain the expression of pressure from eq.(\ref{eueos}) using eq. (\ref{eq5}),
\begin{equation}\label{eq6}
p = -\frac{B^{2}}{(1+A)^{2}}+\frac{BK(A-1)}{(1+A)^{2}}\frac{1}{a^{\frac{3}{2}(1+A)}} +\frac{AK^{2}}{(1+A)^{2}}\frac{1}{a^{3(1+A)}},
\end{equation}
where we have identified different barotropic fluids  as follows: $p_{1}=-\frac{B^{2}}{(1+A)^{2}}$, $p_{2}=\frac{BK(A-1)}{(1+A)^{2}}\frac{1}{a^{\frac{3}{2}(1+A)}}$ and $p_{3}=\frac{AK^{2}}{(1+A)^{2}}\frac{1}{a^{3(1+A)}}$ respectively.
The energy density in the case of an EU is a composition of three different fluid components \cite{Mukherjee_2006}. The first term can be interpreted as a cosmological constant that accommodates the DE sector of the universe. Comparing the above equation with the barotropic EoS $p_{i}=\omega_{i}\rho_{i}$ (where $i = 1,2,3$), with $\omega_{i}$ being the EoS parameter for the ${\it i^{th}}$ fluid, we can obtain the EoS parameters for the individual fluids as $\omega_{1} = -1$, $\omega_{2} = \frac{A-1}{2}$ and $\omega_{3} = A$. The composition of the cosmic fluid depends on the value of the parameter $A$ as determined by Mukherjee et al. \cite{Mukherjee_2006}, e.g. for $A=\frac{1}{3}$ the EU is composed of three types of fluids, dark energy ($\omega_{1}=-1$), cosmic string ($\omega_{2}=-\frac{1}{3}$) and radiation ($\omega_{3}=\frac{1}{3})$ admitting a non-singular model given by eq. \ref{a} and $A=0$ leads to DE ($\omega_{1}=-1$), exotic matter ($\omega_{2}=-\frac{1}{2}$), and dust ($\omega_{3}=0)$. So for a specific value of $A$, the composition of the cosmic fluid is fixed. It is further shown by Paul and Majumdar \cite{Paul_2015} that even if one begins with a given $A$, fluid composition transforms into different types when an interaction sets in that depends on the strength of interaction at the later epoch. In the next section, we consider an interacting fluid scenario for exploring the evolution of the EU. Now for analyzing the model with the observations it is important to represent the scale factor relation with the redshift parameter $z$ given by  $a=\frac{1}{(1+z)}$, where $a(t)$ is the scale factor at any time and we assume the present scale factor of the universe, $a_{0}=1$. The energy density of the universe can be expressed as $\rho=\sum_{i=1}^{3}\rho_{i}$, and can be expressed in terms of $z$ as: $\rho_{1}=\frac{B^{2}}{(1+A)^{2}}$, $\rho_{2}=\frac{2BK}{(1+A)^{2}}(1+z)^{\frac{3}{2}(1+A)}$ and $\rho_{3}=\frac{K^{2}}{(1+A)^{2}}(1+z)^{3(1+A)}$. 
\section{Cosmological models with interacting fluids}\label{intf}
In this section, we study the effect of interaction among the cosmic fluid components. For a given $A = \frac{1}{3}$, the EU is composed of DE, cosmic string, and radiation in the absence of interaction. There are a variety of reasons for the origin of interactions among the cosmic fluids. We assume the interaction among the fluids sets in at $t>t_{i}$, where $t_{i}$ is the time when interaction began. We also assume that there is an interaction between the DE and radiation sectors only while the cosmic string remains non-interacting. The conservation equations for DE ($\rho_{1}$) and radiation ($\rho_{3}$) can be written as \cite{PhysRevD.73.103520, PhysRevD.81.043525, PhysRevD.81.023007, PhysRevD.83.023528},
\begin{equation}\label{eq7}
\dot{\rho}_{1} +3H(\rho_{1} + p_{1}) = - Q,   
\end{equation}
\begin{equation}\label{eq8}
\dot{\rho}_{3} +3H(\rho_{3} + p_{3}) = Q,    
\end{equation}
where, $\rho_{1}$, $p_{1}$ and $\rho_{3}$, $p_{3}$ are the energy densities and pressures of the dark energy and radiation sectors respectively and $Q$ represents the strength of interaction which may assume arbitrary forms. There are no strict constraints on the sign of $Q$ and depending on its sign energy may flow from one sector of fluid to the other. When $Q>0$ energy flows from the dark energy sector to the radiation sector and for $Q<0$ radiation sector loses energy. It is evident from eq. (\ref{eq7}) and eq. (\ref{eq8}) that the individual fluids violate the conservation equation however the total energy density of the fluid remains conserved. The above conservation equations can be recast in the usual form as \cite{Paul_2015},
\begin{equation}\label{eq9}
    \dot{\rho}_{1} + 3H(1 + \omega_{1}^{eff})\rho_{1} = 0
\end{equation}
\begin{equation}\label{eq10}
    \dot{\rho}_{3} + 3H(1 + \omega_{3}^{eff})\rho_{3} = 0
\end{equation}
where $\omega_{1}^{eff}$ and $\omega_{3}^{eff}$ are the effective EoS parameters defined as,
\begin{equation}\label{eq11}
    \omega_{1}^{eff} = \omega_{1} + \frac{Q}{3H\rho_{1}}, 
\end{equation}
\begin{equation}\label{eq12}
    \omega_{3}^{eff} = \omega_{3} - \frac{Q}{3H\rho_{3}}. 
\end{equation}
In the literature, different functional forms of interactions were taken up. There are no strict rules to assume a particular form of interaction however some phenomenological choices are made initially which is then verified using astronomical observations. Several authors have considered different forms of $Q$ such as $Q \propto \rho_{1}$ \cite{PhysRevD.85.043007}, $Q \propto \dot{\rho}_{1}$ \cite{V_liviita_2008}, $Q \propto \rho_{3}$ \cite{10.1111/j.1365-2966.2009.16115.x, di_valentino_can_2017-1}. Cosmological models obtained using several of these interactions are found to be consistent with the observational results \cite{PhysRevD.96.123508, PhysRevD.97.043529}. Thus any new interaction form must be constrained using observations to construct a stable cosmological model. In this paper, we consider a non-linear exponential form of interaction given by,
\begin{equation}\label{eq13}
    Q = 3\;H\;\eta\;e^{(\alpha - 1)},
\end{equation} 
where $\eta$ is a coupling parameter that denotes the interaction strength and $\alpha = \frac{\rho_{1}}{\rho_{3}}$, with $\rho_{i}$ being the energy density of the $i^{th}$ fluid. For $\alpha \rightarrow 1$ the exponential interaction reduces to a linear one. Yang et al. \cite{Yang_2018} obtained observational bounds on the cosmological parameters using such an exponential interaction in the $\Lambda CDM$ model. Recently, Chanda ${\it et. al.}$  \cite{Chandapaul_2023} employed the exponential form of interaction to obtain cosmological models in modified $f(R, \mathcal{G})$ gravity, where $\mathcal{G}$ is the Gauss-Bonnet term. Observational bound on the coupling parameter $\eta$ was obtained using Union 2.1 supernovae data. In both cases, the present observations preferred a small value of $\eta$. In this paper, we construct an interacting EU model and probe the observational viability of the model.\\
The total energy density for the cosmic fluid obtained using Eqs. (\ref{eq5}), (\ref{eq9}) and (\ref{eq10}) yields,
\begin{widetext}
\begin{equation}\label{eq14}
\rho(z) = \rho_{10}(1 + z)^{3(1 +\omega_{1}^{eff})} + \rho_{20}(1 + z)^{2} + \rho_{30}(1 + z)^{3(1 +\omega_{3}^{eff})},
\end{equation}
\end{widetext}
where $\rho_{10} = \frac{B^2}{(1 + A)^2}$, $\rho_{20} = \frac{2BK}{(1 + A)^2}$ and $\rho_{30} = \frac{K^2}{(1 + A)^2}$, and the effective EoS parameters are,
\begin{equation}\label{eq15}
    \omega_{1}^{eff} = -1 +\eta\; e^{(\alpha - 1)},
\end{equation}
\begin{equation}\label{eq16}
    \omega_{3}^{eff} = A - \eta\;\alpha\; e^{(\alpha - 1)}.
\end{equation}
 In the original EU \cite{Mukherjee_2006}, the matter-energy content of the universe is fixed once $A$ is specified and remains so throughout the universe's evolution. However, throughout its evolution, the universe transits from different phases when the matter composition of the universe changes, and different components dominate at different epochs. If one considers an interacting fluid scenario, it is possible to incorporate such transitions at different phases of evolution \cite{Paul_2015}. We note from eq. (\ref{eq16}) that as the strength of the coupling parameter $\eta$ increases the value of $\omega_{3}^{eff}$ decreases approaching zero. Thus for a specific value of $\eta$, $B$, and $K$, the EU transits from a radiation-dominated phase to a matter-dominated phase with an increase in the DE density when,
\begin{equation}\label{effective}
    A = \eta\; \alpha\; e^{(\alpha -1)}. 
\end{equation}
It is also noted that for any $A$ value to begin with (leading to different compositions of matter-energy), the universe always transits into the matter-dominated epoch and gradually evolves into the present observed universe. Thus for a radiation-dominated universe, the value of interaction strength for which the universe transits into a dark energy and matter-dominated one depends on the ratio of the energy densities and is given by,
\begin{equation}\label{eq17}
    \eta = \frac{1}{3\;\alpha  \; e^{(\alpha - 1)}}.
\end{equation}
For a fixed value of $A$, the Friedmann equation (\ref{eq3}) can be expressed in the following form using eq. (\ref{eq14}) as,
\begin{widetext}
\begin{equation}\label{eq17a}
H^{2}(z) = H_{0}^{2}\Big(\Omega_{1}(1 + z)^{3(1+\omega_{1}^{eff})} + \Omega_{2}(1+z)^{\frac{3}{2}} + \Omega_{1}(1 + z)^{3(1+\omega_{3}^{eff})}\Big),
\end{equation}
\end{widetext}
where $\Omega_{i} = \frac{\rho_{i}}{\rho_{c}}$ is the density parameter for the $i^{th}$ fluid, $\rho_{c} = \frac{3H_{0}^{2}}{8\pi G}$ is the critical density and $H_{0} = 100h\;km\; sec^{-1}\; Mpc^{-1}$ is the present day value of the Hubble parameter. For a fixed $\eta$, the values of $B' = \frac{B}{\sqrt{3}H_{0}}$, and $K' = \frac{K}{\sqrt{3}H_{0}}$ for which EU transits from a radiation dominated phase to a DE and matter dominated phase can be obtained by fitting the model with observational data which will be done in the next section.
\section{Constraining the model parameters using observational data}\label{obssec}
This section considers a flat EU with $A=\frac{1}{3}$ and an interaction between the DE and radiation sector only. The model parameters $B'$, and $K'$ are constrained using the Hubble and Pantheon datasets for a specific value of $\alpha$. The Hubble parameter from eq.(\ref{eq17a}) can be represented in the following functional form,
\begin{equation}\label{eq18}
    H^{2}(H_{0},B',K',z) = H_{0}^{2}E^{2}(B',K',z),
\end{equation}
where,
\begin{widetext}
\begin{equation}\label{eq19}
    E^{2}(z) = (\Omega_{1}(1+z)^{3(1+\omega_{1}^{eff})} + \Omega_{2}(1+z)^{\frac{3}{2}} + \Omega_{3}(1+z)^{3(1+\omega_{3}^{eff})}).
\end{equation}
\end{widetext}
In the above equation, $\Omega_{i}$ denotes the density parameter corresponding to the $i^{th}$ fluid where $i=1,2,3$. This expression will be employed to fit the theoretical model with observational data.
\subsection{Observed Hubble Datasets}
The Hubble parameter $H$ can be measured following two different approaches at certain redshifts. The first approach extracts $H(z)$ from the line-of-sight BAO data which includes the correlation functions of the luminous red galaxies and in the second approach $H(z)$ is measured from the differential ages (DA) ($\Delta t$) of the galaxies. In terms of $\Delta t$, the Hubble parameter can be expressed as,
\begin{equation}\label{eq20}
    H(z) = \frac{\dot{a}}{a}=-\frac{1}{1+z}\frac{dz}{dt}\approx -\frac{1}{1+z}\frac{\Delta z}{\Delta t}.
\end{equation}
Recently, Sharov and Vasiliev compiled a list of 57 data points for $H(z)$ in the redshift range $0.07\le z \le 2.42$ \cite{Sharov_2018}. The dataset includes 31 points measured by the DA method (also known as the cosmic chronometer technique) and 26 from BAO and other measurements as shown in Table \ref{tabd}. The $\chi^{2}$ function can be defined as,
\begin{widetext}
\begin{equation}\label{eq21}
    \chi^{2}_{OHD}(H_{0},B',K',z) = \sum_{i=1}^{57}\frac{(H_{th}(H_{0},B',K',z)-H_{obs,i}(z))^{2}}{\sigma_{H,i}^{2}},
\end{equation}
\end{widetext}
where $H_{th}$ is the value of the Hubble parameter estimated from the theoretical model, $H_{obs}(z)$ is the observed Hubble parameter and $\sigma_{H}$ is the standard error associated with the measurement. The present value of the Hubble parameter ($H_{0}$) is treated as a nuisance parameter in this case and its value is taken to be $H_{0}=73.24\pm 1.74$ \cite{riess_cosmic_2021} with a fixed prior distribution for the estimation of $\eta$, $B'$ and $K'$. The parameter $\alpha$ denotes the ratio of the DE density and the matter density and for the present universe must be greater than one. We have assumed $\alpha=2.5$ and $\eta = 0.03$ to obtain a reasonable estimation for the present energy budget of the universe.
\begin{table}
\centering
\caption{$H(z)-z$ dataset with errors estimated from DA and BAO methods}
\begin{tabular}{cccccc}\hline
&       &    Hubble data  &       &\\
\hline
$z$           & $H(z)$  & $\sigma_{H}$    & $z$    & $H(z)$  & $\sigma_{H}$\\
\hline
0.07        & 69    & 19.6    & 0.24 & 79.69 & 2.99  \\
0.90        & 69    & 12      & 0.52 & 94.35 & 2.64  \\
0.120       & 68.6  & 26.2    & 0.3  & 81.7  & 6.22  \\
0.170       & 83    & 8       & 0.56 & 93.34 & 2.3   \\
0.1791      & 75    & 4       & 0.31 & 78.18 & 4.74  \\
0.1993      & 75    & 5       & 0.57 & 87.6  & 7.8   \\
0.200       & 72.9  & 29.6    & 0.34 & 83.8  & 3.66  \\
0.270       & 77    & 14      & 0.57 & 96.8  & 3.4   \\
0.280       & 88.8  & 36.6    & 0.35 & 82.7  & 9.1   \\
0.3519      & 83    & 14      & 0.59 & 98.48 & 3.18  \\
0.3802      & 83    & 13.5    & 0.36 & 79.94 & 3.38  \\
0.400       & 95    & 17      & 0.6  & 87.9  & 6.1   \\
0.4004      & 77    & 10.2    & 0.38 & 81.5  & 1.9   \\
0.4247      & 87.1  & 11.2    & 0.61 & 97.3  & 2.1   \\
0.4497      & 92.8  & 12.9    & 0.4  & 82.04 & 2.03  \\
0.470       & 89    & 34      & 0.64 & 98.82 & 2.98  \\
0.4783      & 80.9  & 9       & 0.43 & 86.45 & 3.97  \\
0.480       & 97    & 62      & 0.73 & 97.3  & 7.0   \\
0.593       & 104   & 13      & 0.44 & 82.6  & 7.8   \\
0.6797      & 92    & 8       & 2.30 & 224   & 8.6   \\
0.7812      & 105   & 12      & 0.44 & 84.81 & 1.83  \\
0.8754      & 125   & 17      & 2.33 & 224   & 8.0   \\
0.880       & 90    & 40      & 0.48 & 87.79 & 2.03  \\
0.900       & 117   & 23      & 2.34 & 222   & 8.5   \\
1.037       & 154   & 20      & 0.51 & 90.4  & 1.9   \\
1.300       & 168   & 17      & 2.36 &  226  & 9.3   \\
1.363       & 160   & 33.6    &      &       &       \\
1.430       & 177   & 18      &      &       &       \\
1.530       & 140   & 14      &      &       &       \\
1.750       & 202   & 40      &      &       &       \\
1.965       & 186.5 & 50.4    &      &       &      \\ \hline
\end{tabular}
\label{tabd}
\end{table}

\begin{figure*}
\centering
\begin{subfigure}{.5\textwidth}
\centering
\includegraphics[width=8cm]{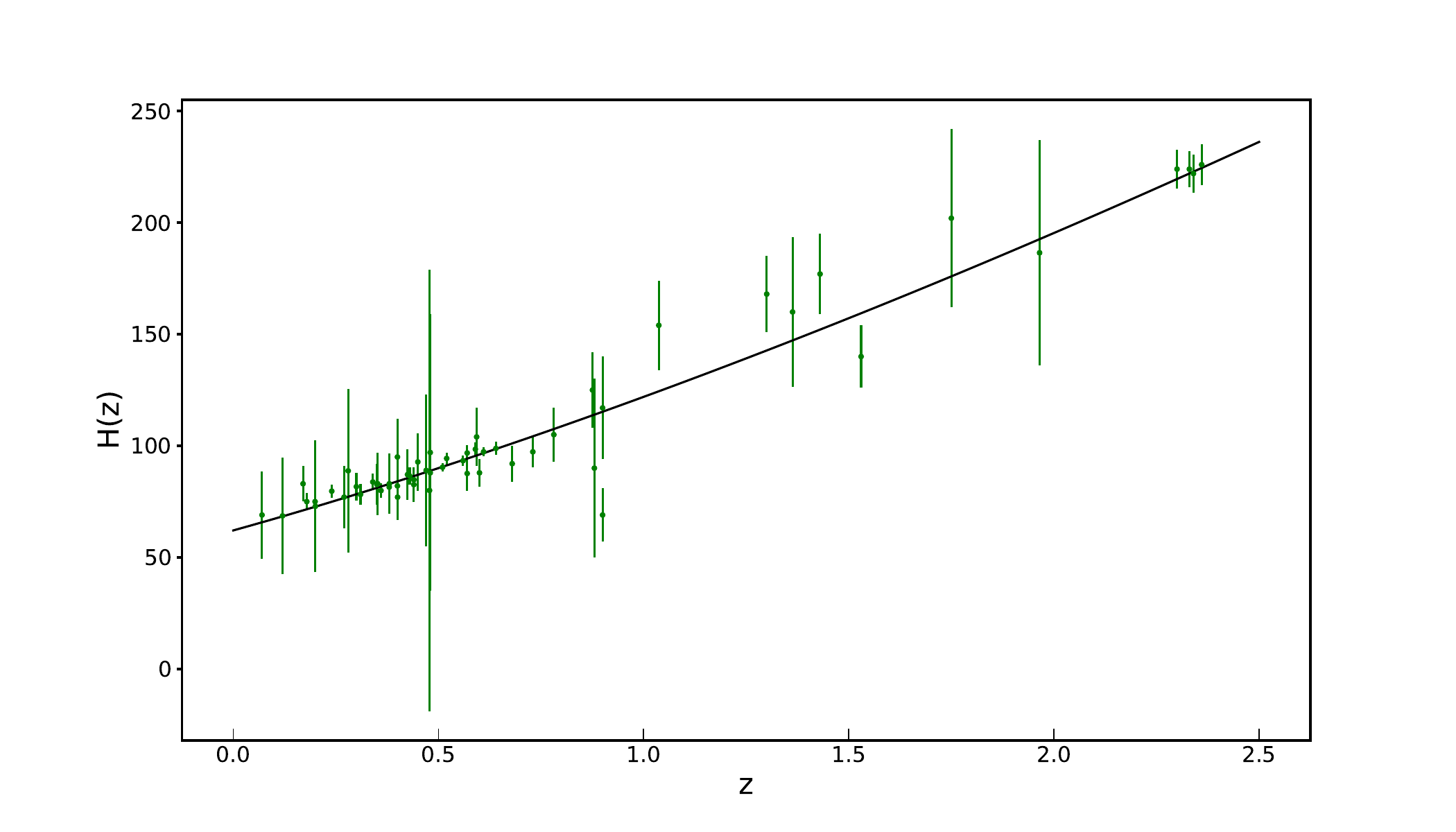}
\caption{}
\label{fit1}
\end{subfigure}%
\begin{subfigure}{.5\textwidth}
\centering
\includegraphics[width=8cm]{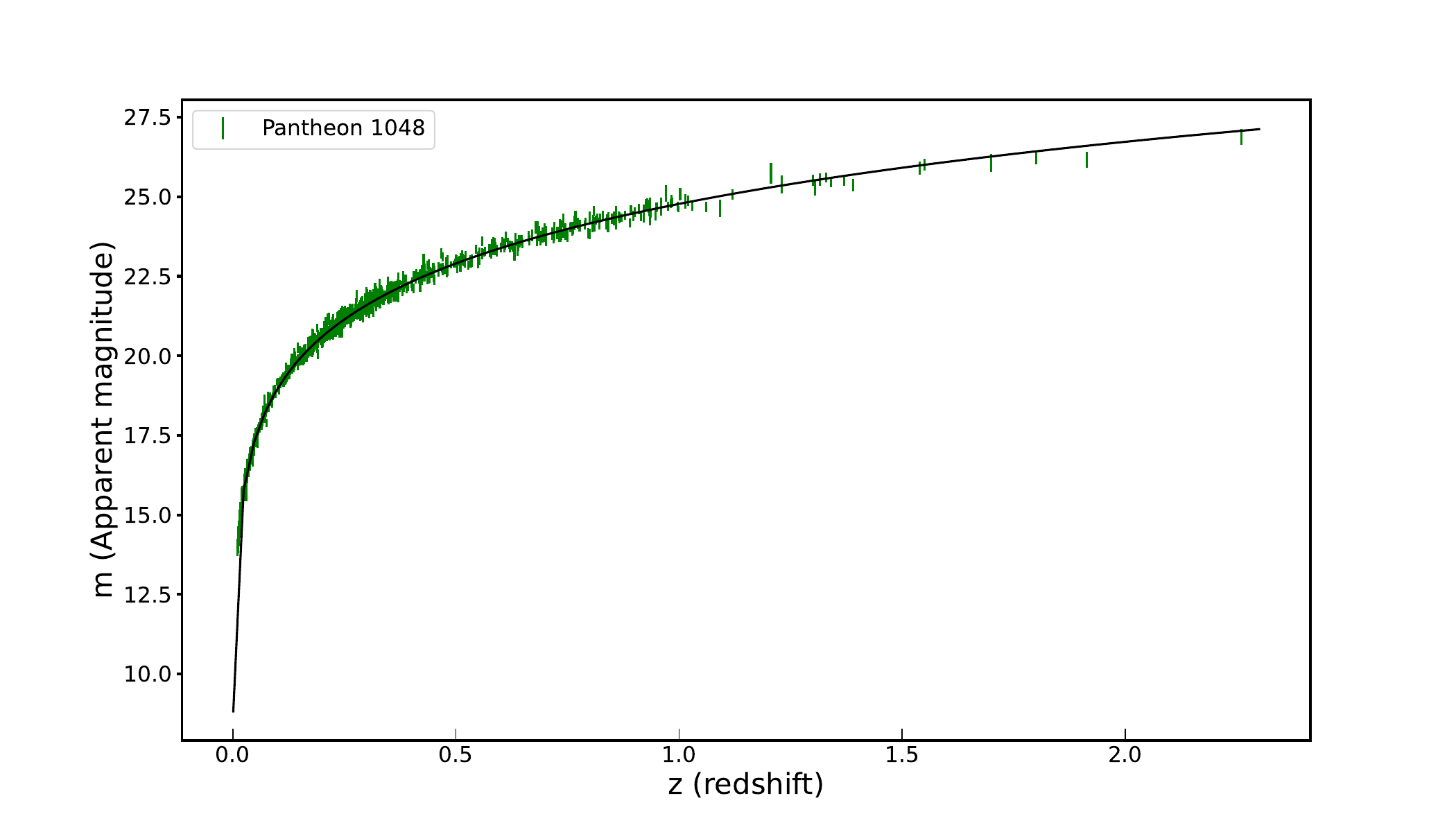}
\caption{}
\label{fit2}
\end{subfigure}%
\caption{The best-fitted curves are shown in the figure. In the left panel, the best-fitted curve corresponding to the 57 Hubble data points is shown with $B'=0.6425$ and $K'=0.4885$, and in the right panel, the corresponding fit for the Pantheon data set is shown with $B'=0.9796$ and $K'=0.3397$ with $\eta = 0.03$.}
\end{figure*}

\begin{figure*}
    \centering
    \includegraphics[width=0.6\textwidth]{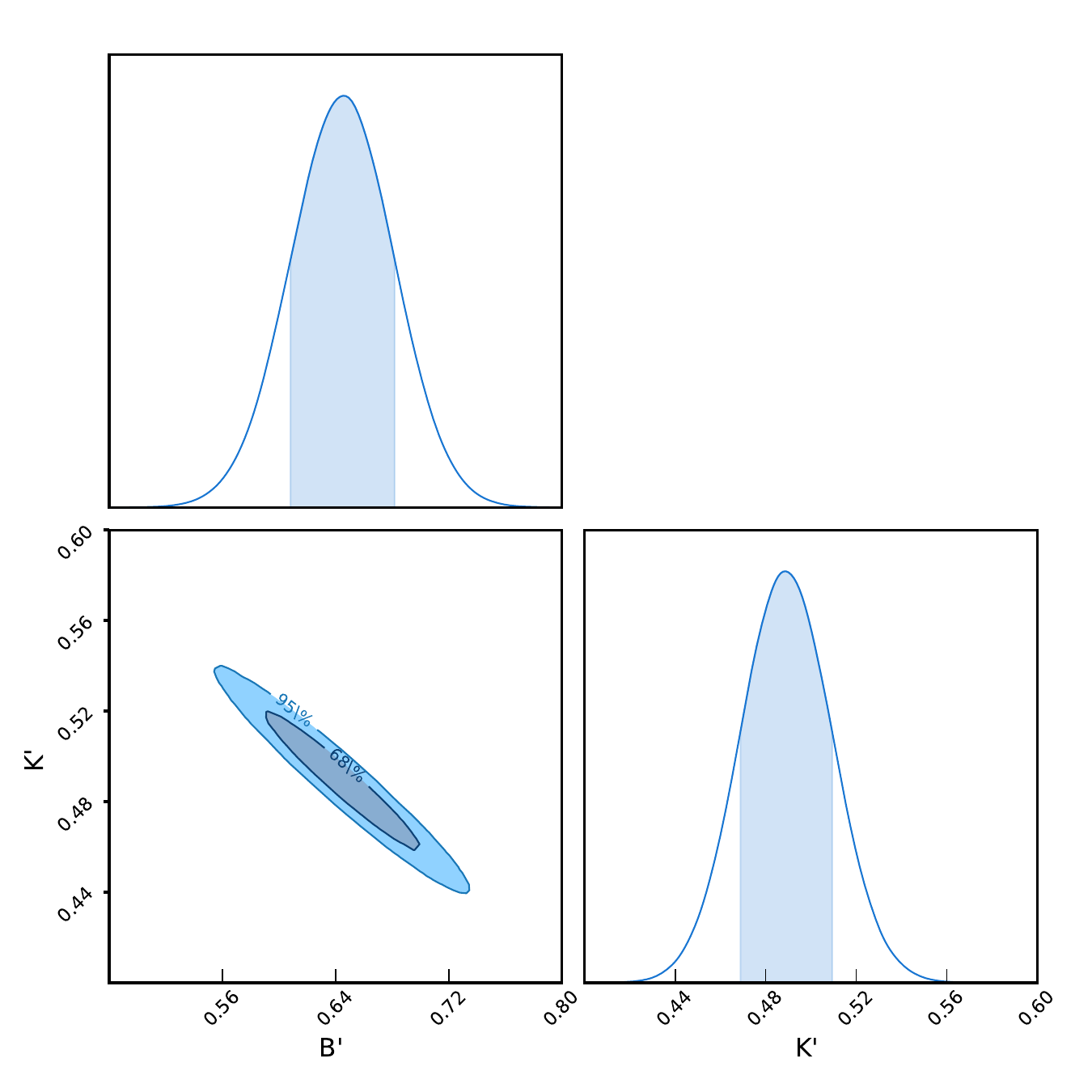}
    \caption{Contours of $1-\sigma$ and $2-\sigma$ confidence levels for the model parameters $B'$ and $K'$ using OHD.}
    \label{ohd}
\end{figure*}

\subsection{Pantheon dataset}
\label{sec:maths} 
The other data set used in the study is the latest Pantheon SnIa sample which consists of spectroscopically confirmed 1048 supernovae specimens compiled by Scolnic et al. \cite{Scolnic_2018}. The sample consists of different supernovae surveys both in the high and low redshift regimes namely, the CfA1-CfA4 surveys \cite{Hicken_2009}, the PanSTARRS1 (PS1) medium deep survey \cite{Scolnic_2018}, the Sloan Digital Sky Survey (SDSS) \cite{Sako_2018}, the SuperNovae Legacy Survey (SNLS) \cite{Guy_2010}, ESSENCE \cite{Narayan_2016}, the Carnegie Supernova project (CSP) \cite{Contreras_2010} and various Hubble space telescope (HST) results \cite{Graur_2014}, \cite{Riess_2018}, \cite{Riess_2007}. For a detailed review and summary of these samples please refer to \cite{Asvesta_2022}. The Pantheon sample covers the redshift range $0.01 < z < 2.26$.\\
The theoretical apparent magnitude of the SnIa can be expressed as,
\begin{equation}\label{eq22}
    m(z) = M + 5\;log_{10}\Big[\frac{d_{L}(z)}{1Mpc}\Big]+25,
\end{equation}
where $M$ is the corrected absolute magnitude. The luminosity distance is denoted by $d_{L}(z)$ and for a flat universe can be expressed as,
\begin{equation}\label{eq23}
    d_{L}(z)=c(1+z)\int_{0}^{z}\frac{dz'}{H(z')},
\end{equation}
with $z$ being the SnIa redshift in the CMB rest frame. One can now define a Hubble free luminosity distance as $D_{L}(z)\equiv \frac{H_{0}d_{L}(z)}{c}$, and the theoretical apparent magnitude in that case becomes,
\begin{equation}\label{eq24}
    m(z)=M+5log_{10}[D_{L}(z)]+5log_{10}\Big(\frac{c/H_{0}}{Mpc}\Big)+25. 
\end{equation}
From the above equation, it is evident that there exists a degeneracy between $M$ and $H_{0}$ and can be combined together to define a new parameter $\mathcal{M}$ as,
\begin{equation}\label{eq25}
    \mathcal{M}=M+5log_{10}\Big[\frac{c/H_{0}}{1Mpc}\Big]+25 = M-5log_{10}(h)+43.28.
\end{equation}
Several attempts have been made to marginalize the degenerate combination and recently \cite{Asvesta_2022} minimized the parameter using the Pantheon sample for a tilted universe. It is seen that the value of $\mathcal{M}$ lies close to 23.8.\\ 
One can now define the $\chi^{2}_{SNS}$ function from the Pantheon sample of 1048 SnIa as,
\begin{equation}\label{eq26}
\chi^{2}_{SNS}(H_{0},\eta,B',K',z) = \Delta \mathcal{F}_{i} C_{SNS}^{-1}\Delta \mathcal{F}_{j},  \end{equation}
where $\Delta \mathcal{F} = \mathcal{F}_{obs} - \mathcal{F}_{th}$ represents the difference between the theoretical and the observed value of the apparent magnitude for each SnIa at redshift $z_{i}$, and $C_{SNS}$ is the total covariance matrix. The total covariance matrix in this case is constructed as a sum of the diagonal matrix containing the statistical uncertainties of the apparent magnitudes (including the photometric error, mass step correction, peculiar velocity and the redshift uncertainty, stochastic gravitational redshift, intrinsic scatter, and distance bias correction) with a non-diagonal matrix constructed from the systematic uncertainties obtained using the bias correction method.

\begin{figure*}[ht]
    \centering
    \includegraphics[width=0.6\textwidth]{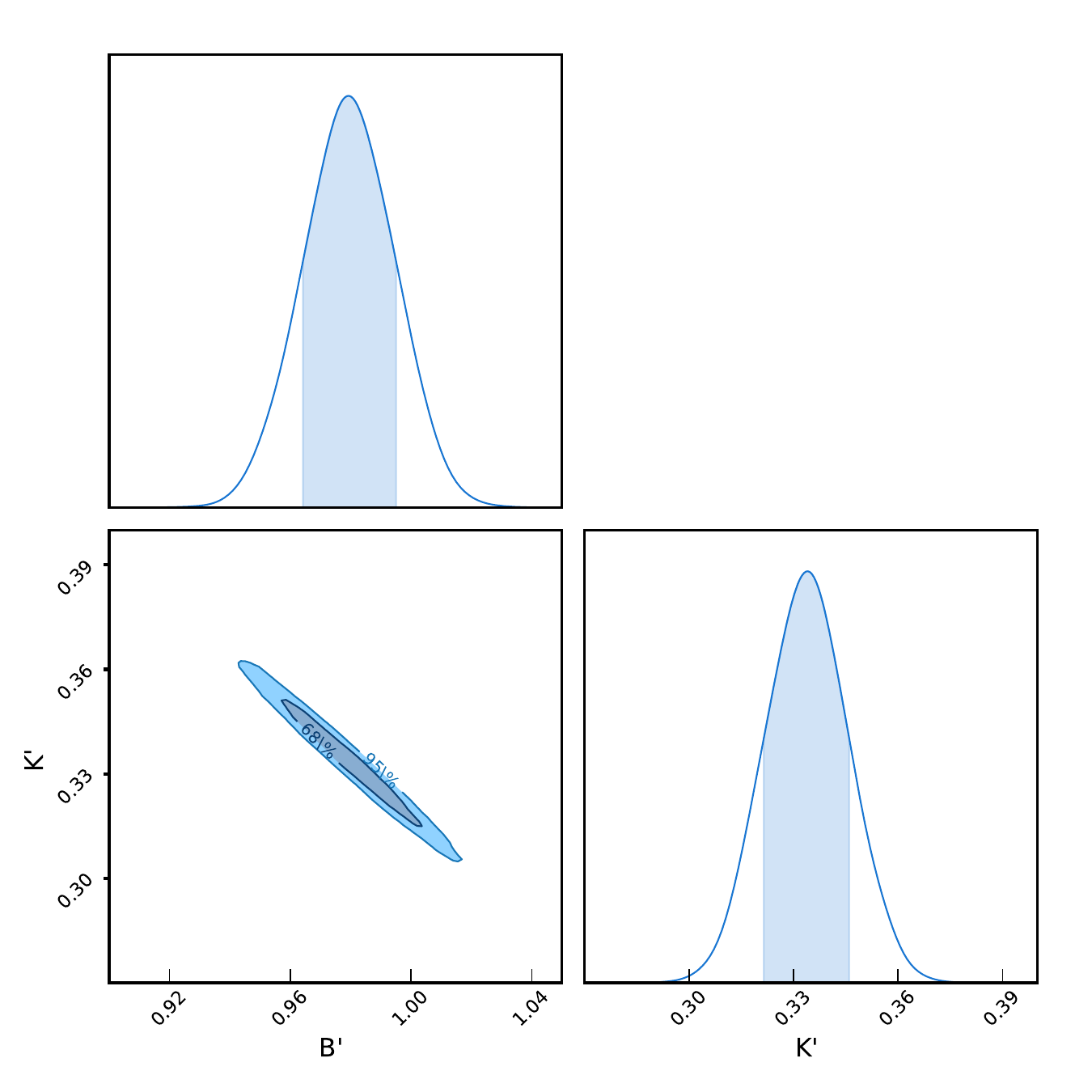}
    \caption{Constraints on the model parameters from the joint analysis of OHD and Pantheon datasets. Contours of $1-\sigma$ and $2-\sigma$ confidence levels for the model parameters $B'$ and $K'$ are shown.}
    \label{joint}
\end{figure*}

We have performed the MCMC analysis to explore the parameter space for the EU model using the python package EMCEE \cite{Foreman-Mackey_2013} and the chains are analyzed using the Chain Consumer \cite{Hinton2016} package which plots the posterior as obtained from the chains. The likelihood function used for the MCMC sampling has the usual functional form,
\begin{equation}\label{eq27}
    \mathcal{L}=\exp(-\frac{\chi^{2}}{2}).
\end{equation}
The best-fitted curves for the OHD and Pantheon data sets with error bars are shown in Figs. \ref{fit1} and \ref{fit2}. To perform the joint analysis with the OHD and Pantheon data we define the joint $\chi^{2}$ function as, $\chi^{2}_{Joint}=\chi^{2}_{OHD} + \chi^{2}_{SNS}$. The joint $\chi^{2}$ function is minimized to obtain the best-fit values for the model parameters. The contours of $1-\sigma$ and $2-\sigma$ confidence level for the parameters $B'$ and $K'$ are shown in Fig. \ref{joint}. The results are summarised in Table \ref{tab1} for OHD and OHD + Pantheon joint analysis. 

\begin{table*}
\centering
\caption{Best fit values of the model parameters}
\resizebox{\textwidth}{!}{ %
\begin{tabular}{|c|cc|cc|}
\hline
        & \;\;\;\;\;\;\;\;\;\;\;\;\;\;OHD        &            & \;\;\;\;\;\;\;\;\;\;\;\;\;\;Pantheon + OHD       & \\
\hline
Parameters & Best fit values  & Mean values $\pm$ $\sigma$ & Best fit values & Mean values $\pm$ $\sigma$\\ \hline
$B'$       & 0.6425    & 0.6425 $\pm$ 0.04     & 0.9796    &  0.979 $\pm$ 0.016    \\ 
$K'$       & 0.4885     & 0.4885 $\pm$ 0.02     & 0.3337    &   0.334 $\pm$ 0.012    \\
\hline
\end{tabular}}
\label{tab1}
\end{table*}

\section{Statistical Inferences with AIC and BIC}
\label{aic}
This section compares the EU model with the standard $\Lambda CDM$ cosmological model using different information criteria. Although there is no particular guideline for the best choice of information criteria, we have used the Akaike Information Criterion (AIC) and Bayesian Information Criterion (BIC) which are quite popular. The AIC is defined as \cite{1100705},
\begin{equation}\label{eq28}
    AIC = \chi^{2}_{min}+2n,
\end{equation}
where $n$ is the number of free parameters in the chosen model. To compare the EU model with the $\Lambda CDM$ model we have used the AIC difference between the two models defined as $\Delta AIC = |AIC_{\Lambda CDM}-AIC_{EU}|$. For two models under consideration, if $\Delta AIC < 2$ then there is strong evidence that the observed data favors the EU model and the model is consistent with the $\Lambda CDM$ model. Whereas, for $4<\Delta AIC \le 7$ there is little evidence in favor of the EU model. If $\Delta AIC>10$ then the model is ruled out \cite{10.1111/j.1745-3933.2007.00306.x}.

The BIC is defined as \cite{10.1214/aos/1176344136}, \cite{Savvas},
\begin{equation}\label{eq29}
    BIC = \chi^{2}_{min}+n\;lnN,
\end{equation}
where $N$ is the number of data points used in the MCMC analysis. It is known that the penalty in BIC is higher than AIC. We denote the BIC difference between the $\Lambda CDM$ and EU model as $\Delta BIC = |BIC_{\Lambda CDM}-BIC_{EU}|$. If $\Delta BIC<2$ then there exists no strong evidence against the EU model as it shows no considerable deviation from the $\Lambda CDM$ model. However, for $2,\Delta BIC<6$ there is evidence against the EU model and for $\Delta BIC>6$, the model is not favored. The difference in the AIC and BIC values and the $\chi^{2}_{min}$ values are displayed in Table \ref{tab2}. We note that $\Delta AIC = 1.6$ and $\Delta BIC = 3.7$ so the model closely resembles the $\Lambda CDM$ cosmology at the present epoch.  

\begin{table*}
\centering
\caption{Goodness of fit $\chi^{2}_{min}$ with the differences $\Delta AIC$ and $\Delta BIC$ for the EU model using the OHD.}
\begin{tabular}{|c|c|c|c|}
\hline
   Model & $\chi^{2}_{min}$ & $\Delta AIC$ & $\Delta BIC$ \\ \hline
   $\Lambda CDM$ &  44.077   & $-$ & $-$\\ \hline
    EU           &  44.439   & 1.6 & 3.7 \\ \hline
\end{tabular}
\label{tab2}
\end{table*}
\section{Cosmological Parameters and Classical Stability}
\label{cosmo}
In this section, we check the viability of the observational constraints by investigating the evolutionary pattern of different cosmological parameters, namely, the deceleration parameter ($q$), the statefinder pair ($r-s$) etc. The deceleration parameter is defined as,
\begin{equation}\label{eq30}
    q = -1 - \frac{\dot{H}}{H^{2}}.
\end{equation}
The deceleration parameter depends on the derivative of the Hubble parameter ($H$). 
\begin{figure*}[ht]
    \centering
    \includegraphics[width=0.6\textwidth]{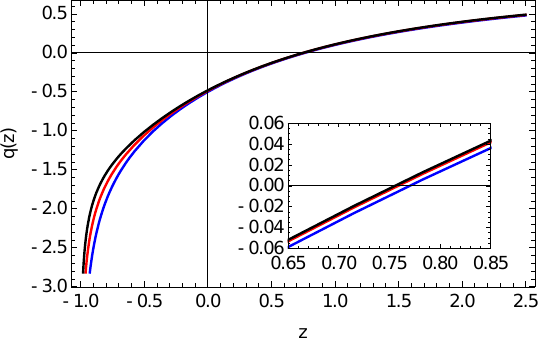}
    \caption{Variation of the deceleration parameter with redshift ($z$) using the best fit values from joint MCMC analysis. The Blue, Red and Black line correspond to $\eta=0.06,0.03$ and $0.015$ respectively.}
    \label{q}
\end{figure*}
In Fig. \ref{q} we have shown the variation of $q$ with redshift ($z$) using the best fit values from joint MCMC analysis using both OHD and Pantheon dataset. From the figure, it is evident that the universe transits from a decelerated phase in the past to an accelerated phase. The present universe is accelerating and it remains in that phase in the near future. The transition redshift, $i.e.$ the redshift at which the universe transits from a decelerated phase of expansion to an accelerating phase of expansion depends on the parameter $\eta$. As $\eta$ increases the universe transits into the accelerating phase at a later time.\\
In the literature, different DE models were proposed to explain the present accelerating universe, namely, quintessence scalar field, phantom, tachyon, Chaplygin gas, etc. To differentiate between different DE models quantitatively Sahni ${\it et. al.}$ \cite{sahni_statefindernew_2003} proposed a geometrical analysis called the statefinder diagnostics. The statefinder parameters $(r,s)$ corresponding to different DE models trace out different geometrical trajectories qualitatively. For the $\Lambda CDM$ model the statefinder pair corresponds to $(r,s) = (1,0)$. The parameters $r$ and $s$ are defined as,
\begin{equation}\label{eq31}
    r = \frac{\dddot{a}}{aH^{3}},
\end{equation}
\begin{equation}\label{eq32}
    s = \frac{r-1}{3(q-\frac{1}{2})}.
\end{equation}
\begin{figure*}
    \centering
    \includegraphics[width=1.0\textwidth]{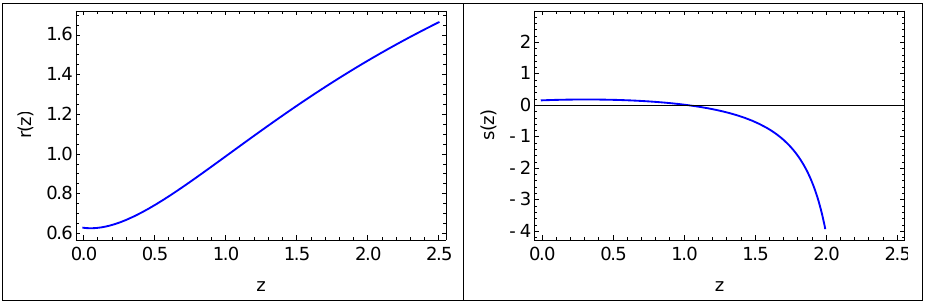}
    \caption{Variation of the statefinder pair ($r,s$) with redshift ($z$) using the bestfit values from joint MCMC analysis.}
    \label{rs}
\end{figure*}
We express the statefinder pair in terms of the deceleration parameter ($q$) as,
\begin{equation}\label{eq33}
    r = q(z)(1+2q(z))+q'(z)(1+z),
\end{equation}
\begin{equation}\label{eq34}
    s(z) = \frac{r(z)-1}{3(q(z)-\frac{1}{2})},
\end{equation}
where "prime"(') denotes the derivative with respect to $z$. For $r<1, s>0$ the model represents the quintessence type of DE whereas for $r>1, s<0$ the model represents Chaplygin gas. We have shown the variation of the statefinder pair with $z$ in Fig. \ref{rs}. From the figure it is evident that initially the EU was filled with CG type of DE. Gradually it made a transition into the quintessence regime passing through the $\Lambda CDM$ phase and at present the universe is quintessence dominated. The change in nature of the DE may be attributed to the interaction which sets in between the cosmic fluid components at some time $t=t_{i}$. This observation is also supported by the fact that the effective EoS for the DE has a value of $\omega_{1}^{eff}=-0.87$ for $\eta = 0.03$. The interaction strength in this case determines the type of DE at the present epoch.

Another important diagnostic tool is the $Om(z)$ diagnostic. The $Om(z)$ parameter is defined as,
\begin{equation}\label{eq35}
    Om(z) = \frac{E^{2}(z)-1}{(1+z)^{3}-1}.
\end{equation}
\begin{figure*}
    \centering
    \includegraphics[width=0.6\textwidth]{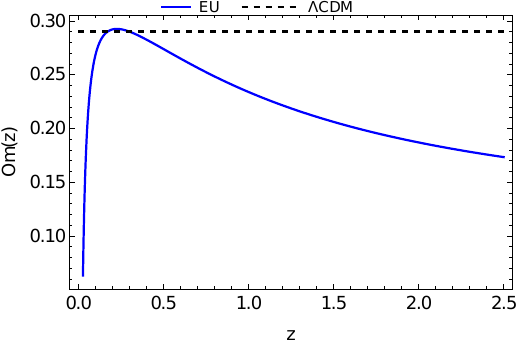}
    \caption{Variation of the $Om(z)$ parameter with redshift ($z$) using the best fit values from joint MCMC analysis.}
    \label{om}
\end{figure*}
The nature of dark energy can be determined by comparing the $Om(z)$ values at two different points. For two different $z$ values, namely $z_{1}$ and $z_{2}$ where $z_{1}<z_{2}$, if $Om(z_{1},z_{2})\equiv Om(z_{1})-Om(z_{2})=0$ then it represents the $\Lambda CDM$ model. For, $Om(z_{1},z_{2})>0$ the DE is of quintessence type. From Fig. \ref{om} it is evident that the present universe is dominated by quintessence type DE as confirmed by the statefinder analysis also. \\
We study the classical stability of the EU scenario against perturbations using the adiabatic sound speed ($c_{s}^{2}=\frac{dp}{d\rho}$). The hydrostatic pressure, in this case, is given by eq.(\ref{eueos}), and the corresponding sound speed is,
\begin{equation}\label{sound}
    c_{s}^{2} = \frac{dp}{d\rho} = A-\frac{B}{2\sqrt{\rho}},
\end{equation}
where $\rho$ is the energy density given by eq. (\ref{eq14}). For a stable cosmological model $0<c_{s}^{2}<1$. Thus from eq. (\ref{sound}), it is evident that for a stable cosmological model $B>0$ and $A>\frac{B}{2\sqrt{\rho}}$. We plot the variation of adiabatic sound against redshift ($z$) in Fig. (\ref{sound}). It is evident that the value of $c_{s}^{2}$ is positive for a theoretically predicted set of values $B'=0.5$ and $K'=0.5$. However, corresponding to the bestfit values obtained using the OHD (Table(\ref{tab1})), $c_{s}^{2}$ is found to flip its sign from positive to negative in the recent past and stays negative at the present epoch ($z=0$). Thus for the observationally predicted values of the model parameters, the EU model exhibits an instability against small perturbations. The small perturbations present in the system will gradually grow in time making the model unstable at the present epoch and near future. In this regard, the stability of various DE models against perturbations is worth looking at. It is found that the Chaplygin gas models and Tachyon models of DE remain stable against small perturbations \cite{PhysRevD.72.103518, PhysRevD.69.123524}. However, several holographic dark energy models with future event horizon is found to be classically unstable throughout the evolutionary history of the universe or in some cases remains stable in the past or future showing instability at the present epoch \cite{MYUNG2007223,SAHA2021168403}. A similar result is obtained for agegraphic DE models with interacting cosmic fluids in the case of both flat and non-flat geometries \cite{KIM2008118}. Recently, Ebrahimi and Sheykhi \cite{ebrahimi_instability_2011} studied the stability of QCD-motivated ghost DE models \cite{PhysRevD.88.043008} using the square speed of sound as the determining factor. It is also found that the cosmological model remains unstable throughout for flat or non-flat geometries even in the presence of interaction between DM and DE.\\
In the present work, we study the stability of an interacting EU scenario where the universe transits from an early radiation-dominated phase (determined by the nEoS parameter $A$) to a matter and DE-dominated phase. We note that although theoretically, it is possible to construct an EU model that remains stable against small perturbations, the observational bounds on the model parameters lead to an EU scenario where the model becomes unstable at the present epoch. The role of the interaction strength $\eta$ is nominal in this case and will be investigated in detail elsewhere.
\begin{figure*}
    \centering
    \includegraphics[width=0.6\textwidth]{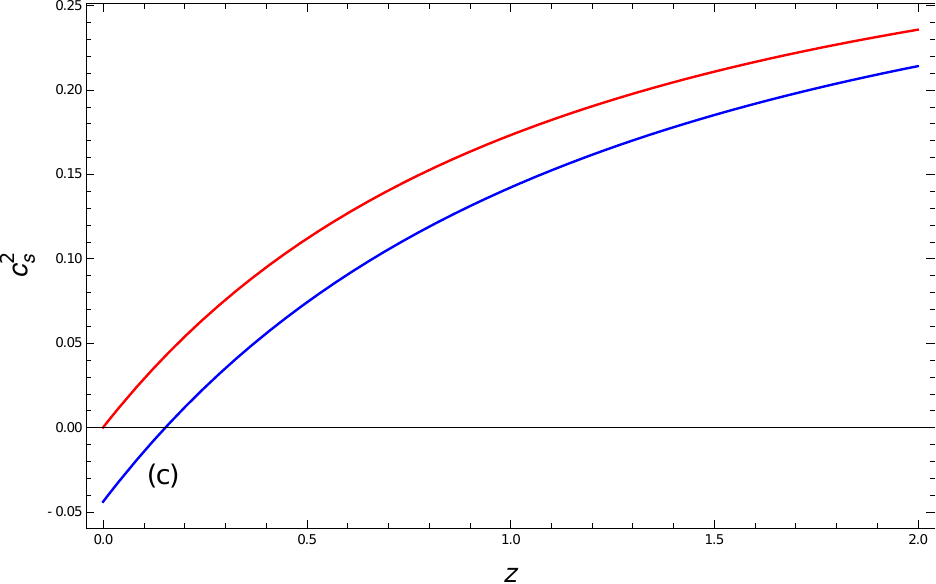}
    \caption{Evolution of square sound speed ($c_{s}^{2}$) against redshift ($z$) for two different set of values of $B'$ and $K'$ with $\alpha=2.5$ and $\eta=0.03$. The red curve corresponds to a theoretical prediction of $B'=0.5$ and $K'=0.5$ and the blue curve corresponds to the $B'$ and $K'$ values as obtained using the OHD.}
    \label{sound}
\end{figure*}

\section{Results and Discussions}
\label{res}
\begin{figure*}
    \centering
    \includegraphics[width=0.6\textwidth]{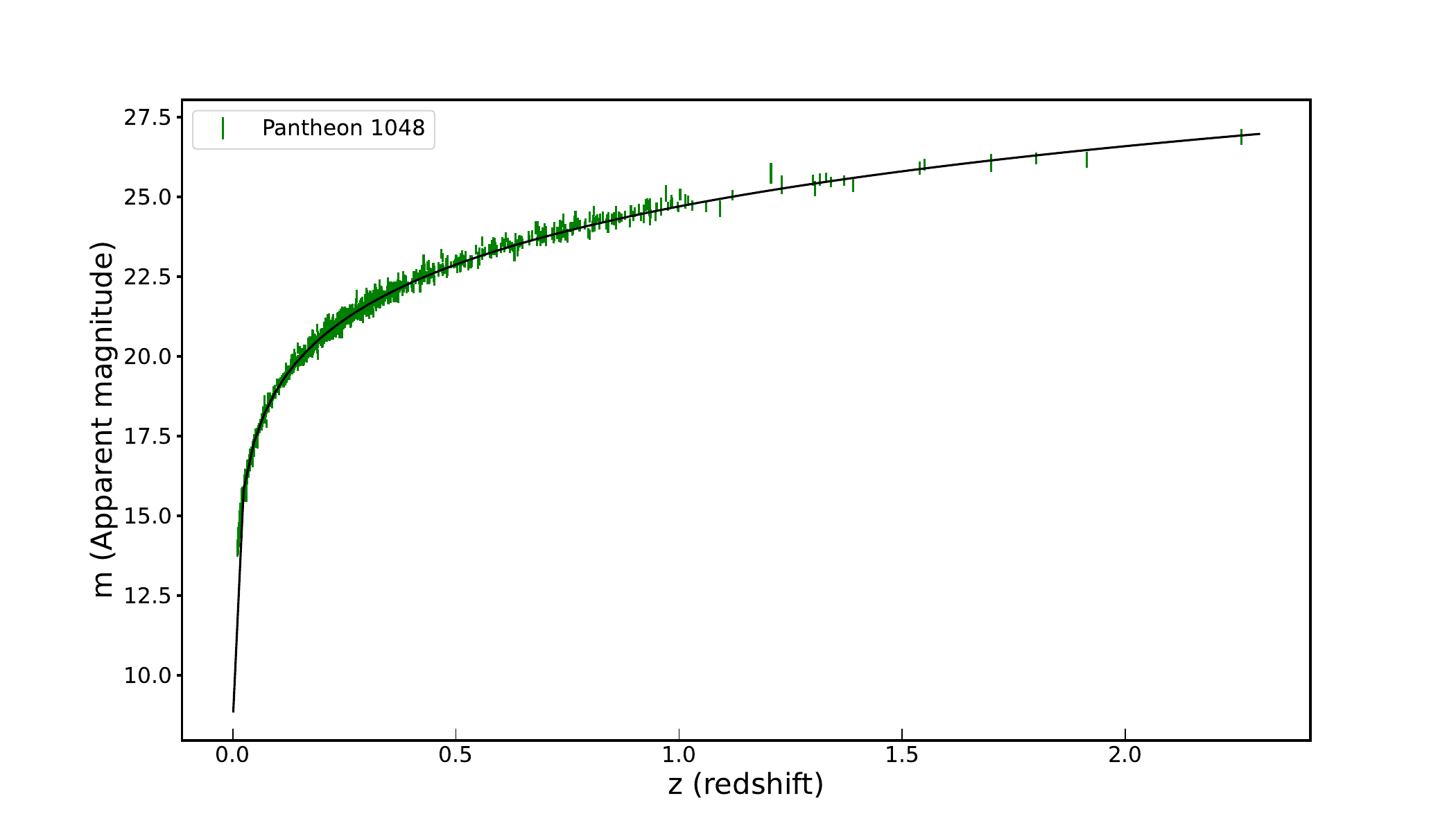}
    \caption{The best-fitted curve for the Pantheon data set is shown with $B' = 0.8315$, $K' = 0.4620$ and $h = 0.67$ with $\eta = 0.03$.}
    \label{3params}
\end{figure*}

\begin{figure*}
    \centering
    \includegraphics[width=0.6\textwidth]{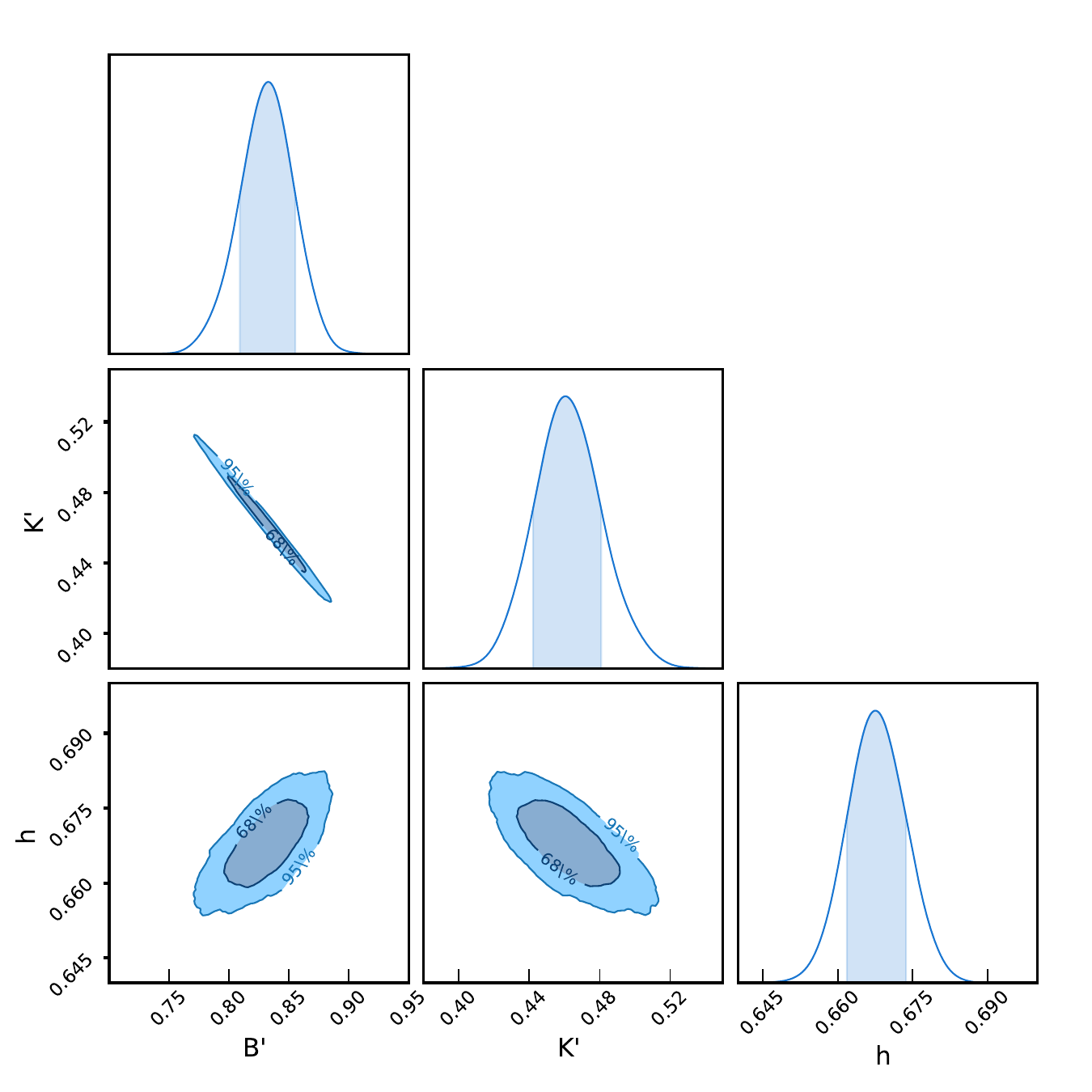}
    \caption{Constraints on the model parameters from the joint analysis of OHD and Pantheon datasets. Contours of $1-\sigma$ and $2-\sigma$ confidence levels for the model parameters $B'$, $K'$ and $h$ are shown.}
    \label{3paramsc}
\end{figure*}
In the paper, EU model obtained by \cite{Mukherjee_2006} is explored in the presence of interaction to estimate the observational constraints on the model parameters and to study the classical stability of the model. It is known that the EU scenario promises to solve some of the well-known conceptual issues in Big Bang theory including technical issues not understood in the standard model. The nEoS of the cosmic fluid is given by eq. (\ref{eueos}) where $A$ and $B$ are the model parameters. It is interesting to note that the nEoS is equivalent to three different fluids as described by eq. (\ref{eq5}). The type of fluids spanning the universe depends on the parameter $A$. In the original EU scenario, once the value of $A$ is fixed, the constituents of the cosmic fluid are determined. However, for a consistent cosmological model, the universe must pass from a radiation-dominated phase to a matter-dominated one and subsequently evolve into the DE-dominated phase. This issue can however be alleviated if an interaction is assumed between the cosmic fluid components. In our analysis, we begin with a specific value of $A$ ($A=\frac{1}{3}$) which leads to an early universe with DE, cosmic string, and radiation. It is shown that an EU can emerge out from the throat of a dynamical wormhole \cite{paul_emergent_2021} which at a later epoch gives rise to a phase with radiation domination (although an insignificant contribution of DE and cosmic string is present) as described by the given nEoS. A non-linear interaction is introduced in the theory to explore the further evolution of the universe. It is noted from the analysis that as the energy exchange takes place between the fluid components the universe effectively transits from a radiation-dominated early universe to a matter and DE-dominated phase in the late time depending upon the strength of interaction. The matter sector in this case contains both baryonic matter as well as CDM. For a specific value of the interaction parameter $\eta$ we constrain the model parameters $B'=\frac{B}{\sqrt{3}H_{0}}$ and $K'=\frac{K}{\sqrt{3}H_{0}}$ using the recent observational data namely the Hubble data which contains the cosmic chronometer as well as the BAO data and the Pantheon SnIa dataset. The model parameters must satisfy the condition $B'>0$ and $K'>0$ for a physically realistic cosmology. We have considered a flat prior for the parameters $B'$ and $K'$ in the range $0 < B' < 2.5$ and $0 < K' < 2.5$ with the value of the Hubble parameter $H_{0}=73.24 \pm 1.74$. In Figs. \ref{fit1} and \ref{fit2}, the bestfit curves for the OHD and Pantheon dataset are shown for the EU model with interaction strength $\eta = 0.03$. The bestfit values are implemented for the MCMC analysis. In Figs. \ref{ohd} and \ref{joint}, the $1-\sigma$ and $2-\sigma$ confidence level contours for $B'$ and $K'$ are shown using the OHD and the joint OHD + Pantheon dataset. We note that in the case of joint analysis, the value of $B'$ is increased from that of the OHD analysis, however, the value of $K'$ decreases in this case. The best fit values of $B'$ and $K'$ and their mean values are displayed in Table \ref{tab1}. The statistical estimations for the EU model are performed following the AIC and BIC and it is seen that the model is consistent with the observations. From the AIC analysis, it is clear that there is strong evidence in favor of EU. The cosmological parameters namely the deceleration parameter ($q$), statefinder pair ($r,s$) and the $Om$ parameter are also obtained using the parameter values obtained from the MCMC analysis. It is evident from Fig. \ref{q} that the universe transits from a decelerated phase of expansion to an accelerated one at some time in the past and remains accelerating in the near future. The transition redshift depends on the interaction strength $\eta$ and for higher $\eta$ values the universe transits into the accelerating phase at a later time as compared to small $\eta$ values. The statefinder and $Om$ diagnostic pathology applied here indicate that the DE began in the form of CG and gradually it evolves and drips away to the quintessence domain crossing the $\Lambda CDM$ regime as shown in Figs. \ref{rs} and \ref{om}, which is a new result. It is noted that the present DE is quintessence type for $\eta=0.03$.\\
The classical stability of the EU model is investigated here using the expression obtained from the square speed of sound. We note that the EU model remains stable for some theoretically predicted values of the model parameters $B'$ and $K'$ at $z=0$. However, if one considers the observational bounds on those parameters then the model becomes unstable against small perturbations near $z=0$ for the chosen strength of interaction $\eta$. Variation of $\eta$ does not change the results to a significant amount. This issue is interesting and requires to be investigated further which will be taken up in future.\\ 
Considering $H_{0}$ as a free parameter along with $B'$ and $K'$ we have performed the joint MCMC analysis with the OHD + Pantheon dataset. The present day value of the Hubble parameter is estimated to be $H_{0}= 67.7 \pm 0.59$ which is close to the results obtained by Chen and Ratra. The best-fitted curve for the Pantheon dataset is shown in Fig. \ref{3params} and the corresponding $1-\sigma$, $2-\sigma$ contours are shown in Fig. \ref{3paramsc}. The interaction strength plays a significant role in determining the value of $H_{0}$ and we note that as $\eta$ decreases the value of $H_{0}$ increases by a small amount.\\
To conclude, we note that even if the EU begins with a specific composition of cosmic fluids, later on in the course of its evolution the universe transits into a new phase of evolution with a set of compositions decided by the strength of interaction among the interacting fluids. It is shown that an interacting EU transforms to a matter-dominated phase with DE resembling the present observed universe. The observational constraints on the model parameters are estimated in addition to the other features of the observed universe with MCMC. The constraints obtained using the OHD and Pantheon leads to a classical instability at the present epoch.

\section*{Acknowledgement}
BCP would like to thank DST SERB for a project (F. No. CRG/2021/000183). AC, BCR and BCP would like to thank IUCAA Centre for Astronomy Research and Development (ICARD), NBU, for extending research facilities. BCR also acknowledges the Ministry of Social Justice and Empowerment, Govt. of India and the University Grants Commission (UGC), India for providing fellowship. The work of KB was partially supported by the JSPS KAKENHI Grant Number 21K03547.

\section*{Data Availability}
There is no new data associated with this article.
\vspace{1cm}


\bibliography{anirban23}

\end{document}